%% file: LBV.tex
\documentclass[12pt,preprint]{aastex}

\slugcomment{To appear in the Astrophysical Journal}
\shorttitle{LBVs}
\shortauthors{Humphreys et al. }

\begin{document}

\title{On the Social Traits of Luminous Blue Variables}

\author{
Roberta M. Humphreys\altaffilmark{1}, Kerstin Weis\altaffilmark{2}, Kris Davidson\altaffilmark{1} and Michael S. Gordon\altaffilmark{1}  
}

\altaffiltext{1}
{Minnesota Institute for Astrophysics, 116 Church St SE, University of Minnesota, Minneapolis, MN 55455; roberta@umn.edu} 

\altaffiltext{2}
{Astronomical Institute, Ruhr-Universitaet Bochum, Germany; 
kweis@astro.rub.de}

\begin{abstract}
In a recent paper, \citet{Smith15} state that the Luminous Blue Variables (LBVs
) in the Milky Way and the Magellanic Clouds are isolated; that they are not 
spatially associated with young  O-type stars.  
They propose a novel explanation that would overturn the standard view of LBVs.  In this paper we test their hypothesis for the LBVs in M31 and M33 as well as the LMC and SMC.  
We show that in M31 and M33,  the LBVs  are  
  associated with luminous young stars and
 supergiants appropriate to their luminosities and positions on the HR Diagram. 
Moreover, in the Smith and Tombleson  scenario most  of the LBVs should be runaway 
 stars,  but the stars' velocities are consistent with  their 
 positions in the respective galaxies.  
 In the Magellanic Clouds,  those authors'  sample was a mixed population. We reassess their
   analysis, removing seven stars that have no clear 
   relation to LBVs. When we separate the more massive {\it classical}  and the 
   {\it less luminous} LBVs,  the classical LBVs have a distribution 
   similar to the late O-type stars,  while the less luminous LBVs have 
   a distribution like the red 
 supergiants. None of the confirmed LBVs have high velocities or are candidate 
 runaway stars. These results support the accepted description of LBVs as 
 evolved massive stars that have shed a lot of mass, and
	     are now close to their Eddington limit. 
\end{abstract} 



\section{Introduction -- Crucial Distinctions Among LBVs} 

Luminous Blue Variables (LBVs) or S Doradus variables  have attracted attention in recent 
years for two mutually independent reasons.   Normal hot-star winds cannot 
account for sufficient mass loss \citep{Fullerton06}, thus another form of
mass loss, perhaps in LBV events, 
is required for them to become WR stars, see also Humphreys \& Davidson (1994).  Second, the non-terminal supernova 
impostors
resemble extreme LBV outbursts \citep{VanDyk} and certain types of 
supernovae were 
observed to have  experienced prior high mass loss events that have been likened to LBVs. 

Unfortunately, serious confusion has arisen because disparate  
objects are often mixed together and collectively called ``LBVs.''     
In a recent paper, \citet{Smith15} argue    
 that LBVs are not generically associated with young massive stars.  
They suggest that LBVs  gained mass from more massive companions in interacting binary systems and subsequently moved away  from their birth sites 
when the companions became  supernovae. 
This would contradict  traditional
views in which evolved
massive hot stars  experience periods of enhanced mass loss,
in transition to a WR star or a supernova.
Their argument is 
based on their failure to find LBVs closely associated with young  O-type stars
in the Milky Way and the Magellanic Clouds.
As we explain in Section 4.2  however, Smith and Tombleson's 
statistical sample is a mixed population and with at least three physically distinct 
types of objects.  When they are separated,  the statistics agree 
with standard expectations.         
We mention this example because it illustrates 
the need to distinguish between giant eruptions, classical LBVs, 
less-luminous LBVs, LBV candidates, and others such as 
B[e] stars that occupy the same part of the HR diagram.  
In this paper we explore those distinctions.

Since there are numerous misunderstandings in this subject, a careful
summary of the background is useful.
LBV/S Dor variables are  {\it evolved massive stars, close to the 
Eddington limit, with a distinctive
spectroscopic and photometric variability.}
There are two classes with different initial masses and evolutionary histories as 
we explain later.  
In its quiescent or normal state, an LBV spectrum resembles 
a B-type supergiant or Of-type/WN star.  During the ``eruption'' or maximum 
visible light stage, increased mass loss causes the wind to become optically thick, 
sometimes called a pseudo-photosphere, at $T \sim$ 7000-9000 K with an absorption line  spectrum resembling  an 
F-type  supergiant. Since this alters the bolometric correction, 
the visual brightness increases by 1--2 magnitudes while the  total luminosity 
remains approximately  constant (Wolf 1989, Humphreys \& Davidson 1994 and numerous 
early references therein) or may decrease \citep{Groh}. Such an event can 
last for several years or even decades. 

There are two recognized classes of LBV/S Dor variables based on 
their position on the HR Diagram \citep{HD94}.  The {\it classical LBVs,\/} with bolometric 
magnitudes between $-9.7$ and $-11.5$ (log L/L$_{\odot}$ $\gtrsim$ 5.8)
have clearly evolved from very 
massive stars with $M_\mathrm{ZAMS} \; \gtrsim \; 50 \; M_\odot$. 
Their high mass loss prevents them from becoming red supergiants 
\citep{HD79}.   
The {\it less luminous} LBVs with M$_{Bol}$ $\simeq$ -8 to  -9.5 mag 
had initial masses in the range  $\sim$ 25 to  40 M$_{\odot}$ or so,  
and can become red supergiants.     

Some factor must distinguish LBVs from the 
far more numerous ordinary stars with similar $T_\mathrm{eff}$ and $L$, 
and $L/M$ is the most evident parameter.  LBVs have larger $L/M$ ratios 
than other stars in the same part of the HR Diagram.  Their Eddington 
factors $\Gamma \, = \, L/L_\mathrm{Edd}$ are around  
0.5 or possibly higher.\footnote{ 
  For example, P Cyg and AG Car have $\Gamma \approx 0.5$ \citep{Vink02,Vink2012}.  }  
  This is not surprising for the very massive classical LBVs.  But how 
  can the less luminous LBVs have such large $L/M$ ratios?
The simplest explanation is that they have passed through a red supergiant 
stage and moved back to the left in the HR diagram \citep{HD94}.   
\citet{MM} showed that stars in the 22 -- 45 M$_{\odot}$ initial mass range, 
evolving back to warmer temperatures, will pass through the LBV stage. Recent 
evolutionary tracks, with mass loss and rotation \citep{Ekstrom},  show that these stars 
will  have shed about half of their initial mass. Having lost 
 much of their mass,  they are now
 fairly  close to $(L/M)_\mathrm{Edd}$. Although empirical mass estimates are uncertain, the low luminosity LBVs are close to their Eddington limit with $\Gamma \, \approx 0.6$ \citep{Vink2012}. 
 Hence {\it the evolutionary state of 
 less-luminous LBVs is fundamentally different from the classical LBVs.\/} 

To illustrate this, in Figure 1  we show the evolution of $L/M$ for two different masses
representing the two LBV/S Dor classes.   
It shows a simplified Eddington factor, 
${\Gamma}_\mathrm{s} = (L/L_\odot)/(43000 M/M_\odot)$,  
as a function of time in four evolution models reported by  
\citet{Ekstrom} for  rotating and non-rotating 
stars with $M_\mathrm{ZAMS} =$ 32 and 60 $M_\odot$.  The 32 and 
60 $M_\odot$ models have $\log L \sim$ 5.6 and 6.0, respectively, 
when they are in the LBV instability strip (see Figs.\ 2 and 3).  
Three critical evolution points are marked in the figure: 
(1) The end of central H burning, (2) the farthest major excursion 
to the red side of the HR Diagram, and (3) the end of central He 
burning.  For the 60 $M_\odot$ star, 
${\Gamma}_\mathrm{s} \gtrsim 0.5$  as it leaves the main sequence, 
but the 32 $M_\odot$ star must pass through yellow and/or red 
supergiant stages before its ${\Gamma}_\mathrm{s}$ reaches 0.5.
Of course, these examples  depend on the  input parameters and 
assumptions of the models.  Other recent evolution models with mass loss
and rotation generically agree on the main point stated here, see for example,
\citet{Chen} 

Therefore we should expect that only the classical LBVs should be associated
with young objects such as O-type stars. The less luminous LBVs, more than 5 million years old, will have moved 50 - 100 pc and are also old enough for their
neighbors to evolve out of the O star class. Among other results in this 
paper, we find that the data agree with this expectation.

\begin{figure}[!h] 
\figurenum{1}
\epsscale{0.5}
\plotone{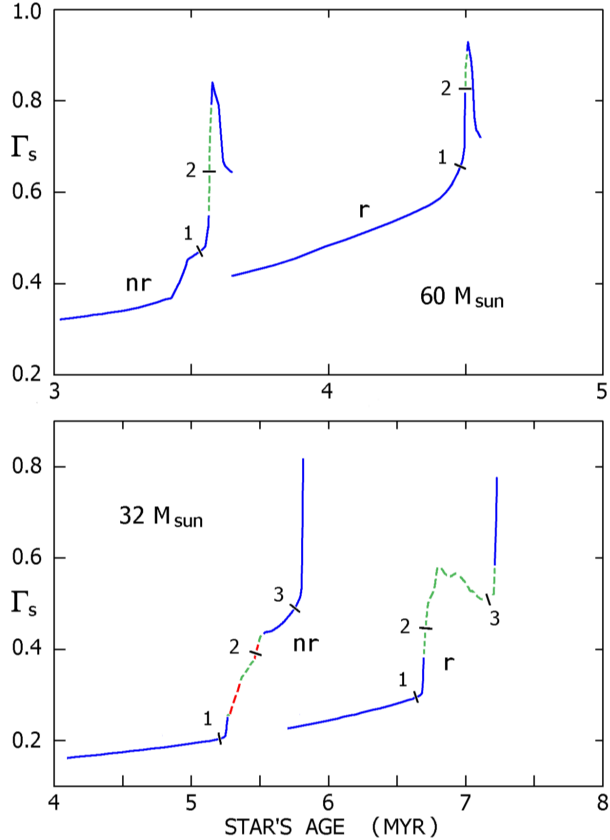}
\caption{The Eddington parameter $\Gamma_\mathrm{s}$ (see text) as a function of
 time for two initial masses, 32 and 60 $M_\odot$, with(r) and without rotation(
 nr). The initial rotation velocities in the models, respectively, are 306 and 346 km s$^{-1}$.  Marked evolution points are (1) end of central H-burning,
 (2) coolest value of $T_{eff}$. and (3) end of central He-burning.
 Blue curves represent stages where $T_{eff} > 10000$ K,
 red indicates $T_{eff} < 4500$ K, and intermediate-temperature
 stages are shown in green.  The dashed segments indicate yellow and red 
 supergiant stages  where low ionization and other factors make ${\Gamma}_s$ 
 ineffective. } 
\end{figure}

One of the distinguishing characteristics of LBV/S Dor variability is that 
during quiescence
or minimum light, the stars  lie on the S Dor instability strip  first introduced by 
\citet{Wolf89} and  illustrated here in Figures ~2 and 3.  
 The more luminous, classical  LBVs 
above the upper luminosity boundary have  not been red supergiants, 
while those below are  post-red supergiant candidates. Thus LBVs with 
very different initial masses and different evolutionary histories occupy the 
same locus in the HR Diagram.  

No single cause is generally
accepted as the origin for the LBV/S Dor enhanced mass loss/optically
thick wind events. Most proposed explanations invoke the star's
proximity to its Eddington limit due to previous high mass loss. Proposed
models include an opacity-modified Eddington limit,
subphotospheric gravity-mode instabilities, super-Eddington winds, and
envelope inflation close to the Eddington limit (see \S {5} in Humphreys \& Davidson 1994, Glatzel 2005, Owocki \& Shaviv 2012, and Vink 2012).

In a {\it giant eruption}, represented by objects like $\eta$ Car
in the 1840's and P Cygni in the 1600's, the star  greatly  increases its
total luminosity with an increase in its visual magnitude typically by
three magnitudes or more.  Giant eruptions should not be confused with
the normal LBV/S Dor variability described above.
 The energetics of the outburst
 and what we observe are  definitely different. Unfortunately, many authors
 do not make the distinction especially with respect to the SN impostors and
 the progenitors of Type IIn supernovae.

Few confirmed LBVs are known in our galaxy due to their rarity, 
uncertainties in distance, and the infrequency of the LBV ``eruption''. 
There are only six confirmed LBVs in the LMC and one in the SMC. Thus, even
the Magellanic Clouds do not provide a large enough sample to confidently 
determine
their relative numbers and group properties relative to other massive star
populations. In this paper we examine the spatial distribution of 
the LBVs and candidate LBVs  in M31 and M33 based on our
recent discussion of the luminous stars in those galaxies which 
included the discovery of a new LBV in M31 \citep{RMH13,RMH14,RMH15}. 
In the next section we show that the majority of these LBVs are found in
or near associations of young stars, and although spectra are not available
for nearby neighbors, their magnitudes and colors support 
the  classification of most of them  as hot supergiants. 

 It must be emphasized  that LBVs are {\it evolved} massive stars. We should not necessarily expect
 to find them closely associated with young O stars. This is especially true
 for the less luminous LBVs. The appropriate
 comparison population should be those supergiants found near
 the quiescent temperature and luminosity of the LBVs on the HR Diagram;
 that is, the evolved massive stars presumably of similar initial mass.
In the Milky Way (Section 3), it is equally important that they be 
at the same distance.
The Smith and Tombleson Magellanic Cloud sample was a mixed population; they combined  the classical LBVs with log L/L$_{\odot}$ $\ge$ -5.8  and 
the less luminous LBVs. In Section 4 we reassess their analysis 
separating the LBVs into the two groups based on their positions 
on the HR Diagram. Their spatial distribution and their kinematics lead 
to significantly  different conclusions
which are summarized in the last section.

\section{The LBVs in M31 and M33} 

The LBVs  in M31 and M33 were
originally known as the Hubble-Sandage variables. In their paper on the 
"brightest variables'', \citet{HS}  indentified one star in M31 (V19 =  AF And)
and four in M33 (Vars. A, B, C, and 2) based on their long term  light
curves from $\approx$ 1920's to 1950 \footnote{Var A is now considered a 
post red supergiant, warm hypergiant \citep{RMH87,RMH06,RMH13}}. 
There are now  five confirmed LBVs in M31: AF And \citep{HS,RMH75}, 
AE And \citep{RMH75}, 
Var 15 \citep{Hubble,RMH78}, Var A-1 \citep{Ros,RMH78}, and the newly 
recognized J004526.62+415006.3 \citep{Shol,RMH15}. The four in M33 are 
 Variables B, C, and 2 \citep{HS,RMH75} and Var 83 \citep{RMH78}. 

These stars plus a few candidate LBVs are listed in Table 1 with their 
luminosities and temperature estimates for their quiescent state  with the 
corresponding references. LBVs have 
one important advantage over other supergiants. During the LBV 
 enhanced mass loss state, with the cool, dense wind, the bolometric 
correction is near zero, allowing us to determine its bolometric luminosity
once corrected for interstellar extinction. This is the case for Var B, and
Var C and for the new LBV, J004526.62+415006.3. Other  
luminosities and temperatures in Table 1 are primarily from \citet{Szeif} or \cite{RMH14}.
The adopted corrections for interstellar extinction are from  \cite{RMH14}.
The LBVs and candidates are shown on an HR Diagram in Figure 2

\input{Table1.tex}

\begin{figure}[!h] 
\figurenum{2}
\epsscale{0.7}
\plotone{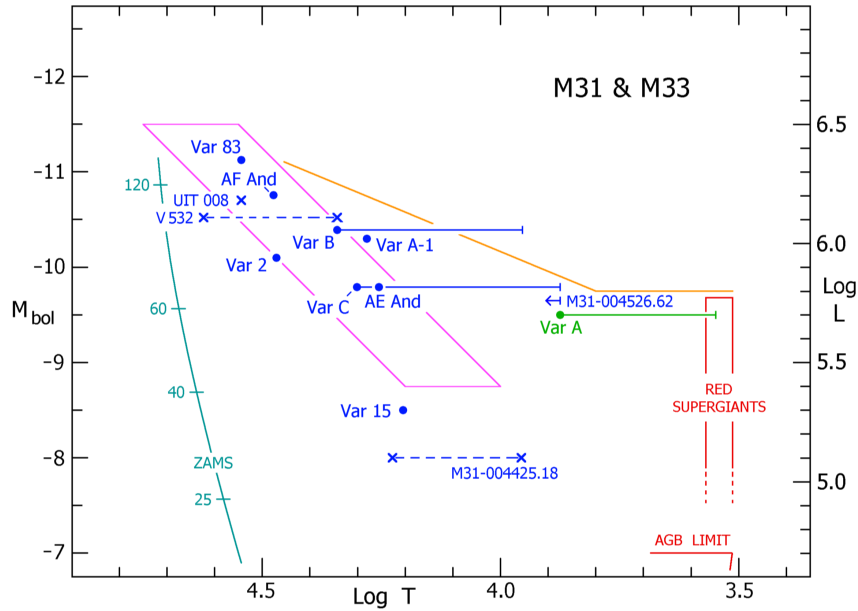}
\caption{A schematic HR Diagram for the LBVs and candidate LBVs in M31 and M33
discussed in the text. The apparent transits in the HR Diagram during the LBV
maximum light or cool, dense wind stage are shown as straight blue lines. The 
LBV/S Dor instability strip is outlined in pink and the empirical upper luminosity
 boundary is shown in orange.}
\end{figure}

Most of the known and candidate LBVs in these galaxies are in known stellar
associations mapped previously by \citet{Hodge} for M31 and by \citet{HS80} in M33 and  a few are also in or near H II regions.  The association designation (A) and number from these references are given in the comments column in Table 1. To compare their space distribution with 
the luminous star populations in these galaxies, we show images in the Appendix of their environments made  from the Local Group  Galaxy Survey \citep{Massey06}.  The LBVs and representative  nearby stars listed in Table 2 are  identified in each image. Although spectra
are not available for these neighboring stars, their magnitudes and colors included 
in Table 2 indicate that most are hot stars or other supergiants.  We use the 
Q--method {\citep{Hiltner,Johnson}} to estimate the intrinsic B-V colors, color excess, and visual extinction (A$_{v}$) with $R = 3.2$  for the candidate OB-type  stars in Table 2. Their absolute 
visual luminosities, M$_{v}$  are determined using distance moduli of 24.4 mag and  24.5 mag for M31 and M33, respectively \citep{M31Ceph,M33Ceph}.

\subsection{Comments on individual stars} 

{\it AE And} had an LBV eruption that lasted for twenty 
years when it was the visually brightest star in M31 \citep{Luyten}. Recent specta show
significant variability in the strengths of the absorption and emission  lines
indicating an unstable  wind. It is in the outer parts of M31 approximately 10{\arcsec} north of A170. Although there 
is no associated H~II region, an arc of emission nebulosity passes through the star (Figure A1). 
Nearby stars include three hot stars and one possible red supergiant. AE And's  luminosity 
 places it just at the upper luminosity boundary, so it could be a post-red supergiant 
 (RSG).  

{\it AF And} is in an association of young stars with a neighboring  
H II region (Figure A1), although this stellar grouping was not included in the 
Hodge catalog. 

{\it Var A-1} is in A42 with an associated H II region (Figure A2). The closest star is a luminous hot star, possibly a late O star, based on its colors.

{\it Var 15} is a ``less luminous'' LBV and therefore  less massive than the other LBVs in M31. It is in A38 and just to the east of a major dust lane (Figure A2). The closest star is red and may be a red supergiant. V15's luminosity and temperature in Table 1 are estimated 
from its current SED \citep{RMH14}. From 1992 to 2001  it got both fainter, by about one magnitude in V, and bluer (Fig. 9 in \citep{RMH14}). The color change suggests that it has gotten hotter.

The new LBV, {\it J004526.62 +415006.3 (M31-004526.62)}, is in A45 with an associated H II region (Figure A4). The three closest stars are all hot stars, probable late O-type supergiants based on their colors and luminosities. 

{\it J004051.59 +403303.0} is a candidate LBV \citep{Massey07,Shol}. It is in the large association A82 
in the outer SE spiral arm (Figure A3). Three of the nearest stars are luminous hot stars. 
Our spectrum from 2013 \citep{GH16} shows prominent asymmetric Balmer emission lines with 
P Cyg profiles and Fe II emission lines with strong P Cyg absorption features. 
The strengths of the Ca II H and K absorption lines and the Mg II 4481 and  He I 4471 lines suggest an early A-type supergiant. Although, it has been called an LBV or LBV candidate it could also be a warm hypergiant \citep{RMH13,GH16}. 

{\it J004425.18 +413452.2} is a potentially interesting star. It shows spectral 
variability reminescent of LBVs, but its luminosity is well below the S Dor instability 
strip \citep{RMH14}. It is in a spiral arm  between the associations A4 and A9 (Figure A3). The nearest stars 
all have intermediate colors typical of yellow supergiants  consistent with 
this star's lower mass and probable  post-RSG status. 

{\it Var B} in M33 has been observed in a recent eruption \citep{Szeif,Massey96}, and consequently has a well-determined luminosity. It is in A142 near the center of M33 (Figure A5). One of the closest objects, marked  {\it d}, 
 may be a compact H II region based on its point-source appearance and very bright H$\alpha$ magnitude. Three additional nearby hot
stars have the colors of late O and early B-type supergiants. 

{\it Var C} has been observed in several maximum light episodes \citep{Burg15} since its initial discovery. 
 It  entered another maximum light phase  in 2013 \citep{RMHVarC}. A period analysis of its light curve suggests possible semi-periodic behavior of 42.4 years \citep{Burg15}.   
Nearby stars include three hot supergiants and a possible RSG (Figure A5). 

{\it Var 83} is a luminous LBV in the prominent southern spiral arm in M33 situated 
between two large associations, A101 and A103. It is surrounded by several prominent nebulous arcs,  and three  nearby neighbors have the colors of O stars and early B-type stars (Figure A6). 

{\it Var 2} is near A100,  10$\arcsec$  above its northern boundary,  and was classified Ofpe/WN9 by \citet{Neugent}. It is surrounded by nebulosity (Figure A6) and the brightest nearby star (a)  
has the colors of an A-type supergiant while  three nearby faint stars have the colors 
of O-type stars and are likely main sequence stars based on their luminosities. 

{\it M33-V532 = GR 290} also known as Romano's star, has been considered an LBV or 
LBV candidate by several authors. Its spectrum  exhibits variability  
from WN8 at minimum to WN11 at visual maximum \citep{Shol2011,Pol2011} with a corresponding  apparent temperature range from  about 42000 K to 22000 K \citep{Shol2011}. But it is 
not known to show the spectroscopic  transition from a hot star  
to the  optically thick cool wind  
at visual maximum that that is  characteristic of the LBV/S Dor phenomenon.
It  may either be in transition to the LBV stage or if it may
be in a post-LBV state \citep{Pol2011,RMH14}. It is included here as a candidate LBV.
It is in the outer parts of M33 about 30$\arcsec$ east of A89 (Figure A7). The nearest star (a) has the colors of an O-type main sequence star.

{\it UIT 008} is another candidate LBV \citep{RMH14}. It has the spectrum of an Of/WN star 
and an unusually slow wind like the LBVs. Its high temperature and luminosity place it
near the top of the S Dor instability strip. It is located in A27 and its   associated H II
region NGC 588 which contains several luminous O stars and WR stars. 

{\it B526 = UIT 341 = M33C-7292}  has been considered an LBV candidate \citep{Clark12}, but it 
is actually two stars
separated by less than 1$\arcsec$. Its published spectra are all likely composite.  
A recent  long slit spectrum observed with the LBT/MODS \citep{RMH16}
clearly separates the two stars 
and shows that the northeast component has an emission line spectrum  often 
associated with LBVs with 
strong H emission with P Cygni profiles plus Fe II and [Fe II] emission. We therefore 
consider this object's northeast  component  a candidate LBV (B526NE). It is in the large 
A101 association with numerous nearby luminous stars (Figure A7). We classify its close 
neighbor to the southwest as a B5 supergiant \citep{RMH16}.

\input{Table2.tex}

\clearpage 

\subsection{Spatial Distribution and Kinematics}

As we have  emphasized, the spatial distribution of the LBVs should 
be compared with a stellar population of similar evolutionary state, with 
 comparable luminosities and initial masses based on their positions on the HR Diagram. 
All of the above luminous LBVs and candidates with M$_{Bol}$ $\gtrsim$ -9.8 mag, are in associations of luminous stars, some with associated H II
regions and emission nebulosity. Their nearby stars in Table 2, for the most 
part have the colors of early-type stars. One might argue that by comparison
AE And looks relatively isolated. With its  luminosity
range, it could be a post-RSG candidate and its association with less
luminous stars would be expected. 
The WR variable, M33-V532  is likewise in the outerparts of M33, 
but near an association with several early type stars.

In the Smith and Tombleson model not only have the LBVs gained mass from a companion,
but they have been ``kicked'' or moved away  from their place of origin after 
the 
supernova explosion of their more massive companion. In that case, we would expect some of the LBVs to have high velocities compared to their expected velocity 
in the parent galaxy. They would be runaway stars.   

Their  radial velocities are in Table 1 together with the expected 
velocity, in parentheses,  at their positions  in M31 and M33. The absorption lines in LBVs show 
considerable variability due to their variable winds. We use the velocities measured from the centroid of [Fe II] and [N II] lines to estimate their systemic velocities. These lines are formed in the low density outer parts of the wind and are not affected 
by P Cygni absorption. The expected velocities were calculated following the prescription and fit to the rotation curves from \citet{Massey09}, \citet{Drout09}, and \citet{Drout12}, for the yellow and red supergiants in M31 and M33.   

Most  of  the LBVs have velocities within 40 km s$^{-1}$  of the expected 
velocity based on the rotation curve. This is consistent with and  
within the velocity range
used by \citet{Drout09,Drout12} to decide on membership of the yellow 
supergiants in M31 and M33.  The velocities for runaway stars depend on the orbital parameters. Based on some of the models, the velocities are expected to be  100 to 200 km s$^{-1}$ and higher \citep{Tauris}. Thus, none of the LBVs or candidates can be described as a runaway star.

\section{The Milky Way LBVs}

There are very few confirmed LBVs in the Milky Way due to the infrequency of
the LBV eruption and of course to the uncertainty in the distances. Smith and
Tombleson 
list 10 LBVs and candidate LBVs and  discuss some of the better studied examples. 
A few deserve some comments and cautionary remarks. As  has been
acknowledged for some time, the giant eruption $\eta$ Car is clearly
associated with a population of luminous massive stars including several
O3-type stars \citep{DH97}. At its presumed initial mass of 150 - 200 M$_{\odot}$, it is at
most about 2 -- 3 million years old and has not have moved far from its place of
birth. 

Smith and Tombleson  emphasize that most of the other Galactic LBVs 
 are relatively isolated from  O stars. But most  are also much 
less massive and will be older.    For example, the well-studied, 
classic stellar wind star,  and survivor of a giant eruption, P Cygni is in the Cyg OB1 
association. At its distance, the two nearest B-type supergiants of similar 
temperature and luminosity are about 30 pc away. The same is true of an O star.
At its position 
on the HR Diagram, P Cyg had an  initial mass of  $\approx$ 50 - 60 M$_{\odot}$ and is past core H-burning with a current mass of 23 to  30 M$_{\odot}$ \citep{Lamers83,PP}.  It is  therefore at least 
4 - 5 $\times$ 10$^{6}$ years old based on models by e.g., \citet{Ekstrom}, with and 
without rotation. With a velocity dispersion of 10 km s$^{-1}$ for the extreme 
Population I, P Cyg   will have moved 35 - 44 pc or more.  
So its spatial separation from  similar stars is what would be expected. 

AG Car is a well-studied classical LBV. At its high luminosity and presumed 
initial mass of $\approx$ 80 M$_{\odot}$, one would expect to find an associated  population
of massive stars.  Smith and Tombleson  argue that there are no nearby O stars. 
However, at its large kinematic distance of $\approx$ 6 kpc \citep{RMH89,Groh}, AG Car  
 must be compared with  the stellar population at the same distance. 
Since our line of sight to AG Car at {\it l} = 289\arcdeg, is directly down 
the Carina spiral arm, the cataloged O stars are most likely in its foreground.

HR Car is in a similar situation at $\sim$ 5 kpc in the Carina arm \citep{vanG91}.
HR Car, HD 160529, and HD 168607  are confirmed LBVs, but belong to the less
luminous, less massive group of LBVs with ages of roughly 6 million years or more; 
enough time for them to move 30 to 100 pc away from their original 
locations even at low random speeds.
We would not expect them  to be closely associated with O stars.

In all cases these stars need to be compared with stars at the same distance and
in the same part of the HR diagram.  
That is of course the advantage of studying these stars in nearby galaxies
and especially in the Magellanic Clouds. 

\section{The LBVs in the Magellanic Clouds}

Smith and Tombleson  included nineteen stars in the  
Magellanic Clouds in their analysis.            
However, six of the  
LMC stars in their list are neither LBVs nor candidates. 
This does not mean that these  stars, described below,  are not interesting or that someday they may be
shown to have LBV characteristics, but they should not be included in
a survey to set limits on the spatial distribution of the known LBVs.
The six confirmed LMC LBVs (S Dor, R71, R127, R143, R110, R85) and four 
candidates (S61, S119, Sk -69\arcdeg 142a, Sk -69\arcdeg 279)  are listed in Table 3  
with their observed parameters.  They also list three stars in the SMC, but R40 is the only confirmed LBV/S Dor variable \citep{Szeifert93}.
The very luminous HD 5980 is a complex triple system that 
may be an example of a giant eruption as we describe below. R4 is a B[e] star. 
The confirmed LBVs are shown on an HR Diagram
in Figure 3. The three most luminous LBVs, R127, S Dor and R143, are 
the classical LBVs \citep{HD94}, and the remaining four belong to the less luminous group.

\begin{figure}[!h] 
\figurenum{3}
\epsscale{0.7}
\plotone{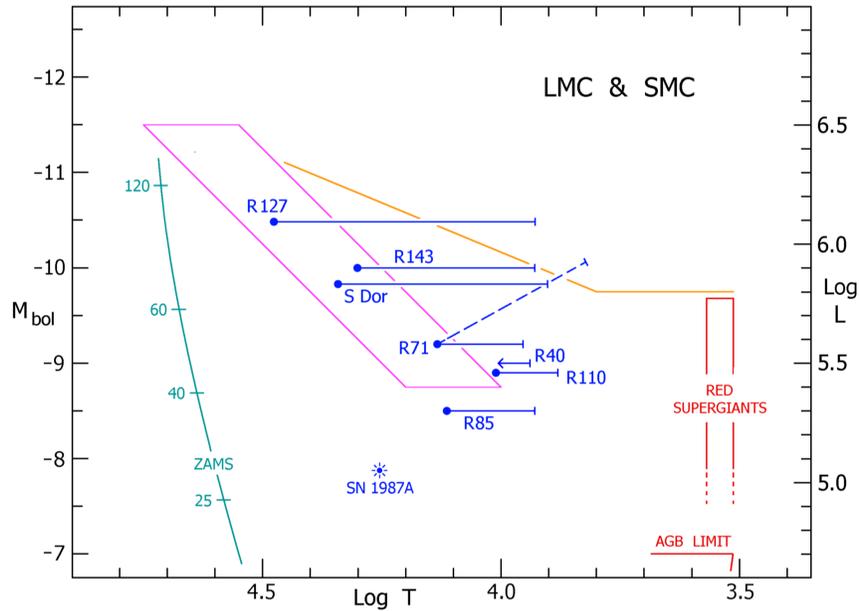}
\caption{A schematic HR Diagram for the LBVs and candidate LBVs in the LMC and S
MC discussed in the text.The symbols and colors are the same as in Figure 2.}
\end{figure}

\input{Table3.tex}

\subsection{The LBV Candidates in the LMC and SMC}

The properties of the  four candidates and  the six excluded stars in the LMC
plus  HD 5980 and the candidate R4 in the SMC are described here with our  reasoning for their inclusion or not in our analysis.

S61 (= Sk -67\arcdeg 266) was  classified  O8Iafpe by \citet{Bohannan} but later revised to WN11h by \citet{Crowther97}. No S Dor 
type variability has been observed and its wind speed of 900 km s$^{-1}$ 
\citep{Wolf87}  is 
much higher than observed for confirmed LBVs in quiescence \citep{RMH14}, although 
\citet{Crowther97} derive a terminal velocity of 250 km s$^{-1}$. S61  
has an expanding circumstellar nebula \citep{Weis2003b} similar to the 
 nebulae associated with some  LBVs.  Primarily for that 
reason it is included as an LBV candidate.

S119 (= HDE 269687) is an Ofpe/WN9 star \citep{Bohannan} or WN11h 
\citep{Crowther97} with the low wind velocity, 230 km s$^{-1}$
typical of LBVs in quiescence, and an expanding  circumstellar nebula 
\citep{Weis2003a}. But \citet{Weis2003a} also find that the center of the 
expansion has a velocity of 156 km s$^{-1}$, rather low for membership in 
the LMC.  The possibility that S119 may be a runaway  is discussed later.
It has not been observed to show  S Dor type 
variability, and is included here as a candidate LBV. 

Sk -69\arcdeg 142a (= HDE 269582) is another late WN star (WN10h, \citet{Crowther97}), a class 
of stars  associated with some LBVs in quiescence. It is a known variable with small light variations on the order of $\pm$ 0.2 mag \citep{vanG}, but
does not have a circumstellar nebula.

Sk -69\arcdeg 279 is classified as an O9f star \citep{Conti86} with a
 large associated circumstellar nebula \citep{Weis95,Weis97,Weis02}.
It is on that basis that we consider it a candidate LBV.

{\it LMC Non-LBVs:}  

R81 (= HDE 269128 = Hen S86) was initially considered an example of 
S Dor variability \citep{Wolf81}, but it was soon shown to be an 
eclipsing binary \citep{Stahl87}.  This analysis and conclusion is
confirmed in the more recent paper by \citet{Tubbesing}. For this reason it was 
not included in the review by \citet{HD94}.  R81's spectroscopic  and photometric
variability are tied to its orbital motion. It is not an LBV or a candidate.  

 MWC 112 (= Sk -69\arcdeg 147) is an F5 Ia supergiant \citep{Rou}. It has not been considered an 
LBV candidate, although it has been confused with HDE 269582 discussed above,
see \citet{vanG,HD94}.  

R126 (= HD 37974) is a B[e]sg star \citep{Lamers,Zickgraf} similar to the 
classic S18, and has not been considered an LBV or candidate LBV in the 
literature. 

R84 (= HDE 269227) has a composite spectrum with an early-type supergiant
plus an M supergiant \citep{Munari}. It has been suggested to be a 
dormant LBV \citep{Crowther1995}, but the various conflicting descriptions
of this star which have included reports of TiO bands, are now 
attributed to its composite nature (B0 Ia + M4 Ia, \citet{Stahl84}).
If this is a physical pair it is an important star since so few M supergiants
are known binaries.

Sk -69\arcdeg 271 (= CPD -69\arcdeg 5001) has been classified as a B4 I/III star
\citep{Neugent11}. It is associated with an arc of H$\alpha$ emission \citep{Weis97} and is considered to be a possible X-ray binary \citep{Sasaki}. It has 
not been suggested to be an LBV or candidate in the literature. Its 
luminosity, based on its spectral type and extinction \citep{Neugent11}, of
M$_{Bol}$ $\approx$ -7.90 mag, places it  below the LBV/S Dor instability 
strip. 

R99 (= HDE 269445) can best be described as peculiar. It is classified 
as an Ofpe/WN9 star \citep{Bohannan}. Although it shows spectral and
photometric variability, the amplitude is very small. \citet{Crowther97}
emphasize its peculiar spectrum and variability. Its high wind velocity of
1000 km s$^{-1}$ argues against an LBV in quiesence. Although it is associated 
with emission from a nearby H II region, \citet{Weis2003b} concludes that there is
no circumstellar nebula  like those associated with some LBVs and candidates.  

{\it A Giant Eruption Candidate and a Non-LBV in the SMC} 

HD 5980 is a complex triple system with two luminous hot  stars in a short
period eclipsing binary (P$_{AB}$ = 19.3$^{d}$). Star B is considered to 
be an early-type WN star, and the third star(C) is an O-type supergiant which may also be a binary \citep{Koenig14}.  
In 1994, HD 5980 experienced a brief 3.5 mag brightening that lasted about 5 months.
At maximum, the absorption line spectrum was described  as B1.5Ia$^{+}$ 
\citep{Koenig}, but \citet{Moffat} later classified it as WN11. It did not produce 
the LBV/S Dor cool wind at maximum. In that respect it is reminiscent of M33-V532. 
Given its brightening and short duration however, 
HD 5980 can be better described as a giant eruption. Nevertheless numerous 
authors describe it as an LBV. For that reason we include it in Table 3, but 
with the above caveat.

R4 is a spectroscopic binary \citep{Zickgraf96} with an early B-type supergiant plus an early A-type supegiant. The hot star is described as a B[e] star with T$_{eff}$ of 27,000\arcdeg and M$_{Bol}$ = -7.7 mag which places it below and well outside
the S Dor instability strip. Its LBV nature has not been established,  
and subsequent papers treat it as a B[e] star.

\subsection{The Spatial Distribution and Kinematics of the Magellanic Cloud LBVs}

Here we reassess the Smith and Tombleson  analysis of spatial 
correlations in the LMC and SMC.   For  each star in a set, e.g. WN stars, red supergiants,  or LBVs, they measured the  projected 
distance $D1$ to the nearest O-type star  and plotted the cumulative 
fraction of the stars in that set that have $D1$  less than a given value.  
This is illustrated in Figure 4 where we reproduce their 
Magellanic sample from  their Table 1  together with the late O-type stars, 
WN stars, and red supergiants. The O-type stars, for example, tend to exist in groups, so naturally they have small values of $D1$. We would conclude from 
this figure that the red supergiants have a spatial distribution uncorrelated 
with O stars. The stars in the Smith and Tombleson  sample fall in between.  

\begin{figure}[!h] 
\figurenum{4} 
\epsscale{0.6}
\plotone{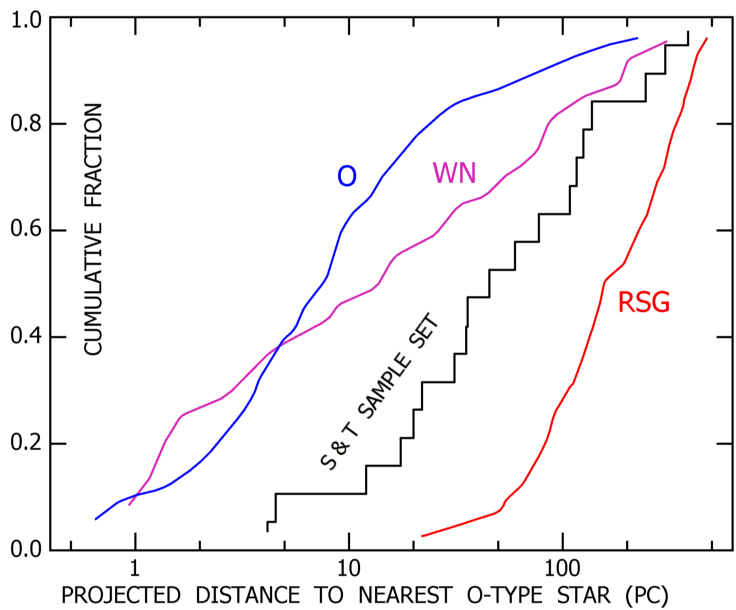}
\caption{The cumulative fraction  versus the projected distance to the nearest O
-type star for the Magellanic Cloud sample from  \citet{Smith15} based on their 
figure. The distributions for the late O-type stars, WN stars, and red supergiants are shown for reference.}
\end{figure}

We repeat this analysis for the seven confirmed LBVs  in Table 3. The 
seven   stars discussed in \S {4.1} that have no clear relation to 
LBVs are removed, and the four  ``candidates'' are discussed below. 
We show the cumulative distributions 
for the {\it classical} and {\it less luminous} LBVs in Figure 5, plotted respectively, as LBV1 and LBV2\footnote{Set LBV1 consists of R127, R143, and S Dor;  
no additional classical LBVs have been found in the LMC and SMC since 1994.  
          Set LBV2 includes R71, R110, R85, and R40.}.  
The three  most luminous stars (LBV1) in the sample were 
originally identified
as classical LBVs by \citet{HD94}, and were not merely chosen in relation to Smith and Tombleson's arguments. They have an average $D1$ of only 7 pc, and,  as 
Figure 5 shows, their distribution is statistically indistinguishable from the late O-type
stars.  The four stars in LBV2  are all substantially fainter than 
  the  $M_\mathrm{bol} \approx -9.7$ criterion.  As expected 
    for these less massive, highly evolved objects, they parallel 
      the RSG distribution in Figure 5.  Formal statistical tests     
        are doubtful for such a small sample;  but if one 
	  attempts to apply any such test (e.g., Kolmogorov-Smirnov),  
	    the result is excellent consistency between set LBV2 and the 
	      RSG distribution.  This is obvious in the figure. 

 Moreover, elementary statistics support the distinction between 
classical and less luminous LBVs.  Among the 7 confirmed LBVs, 
the three classical LBVs have the three smallest values of $D1$. 
The ab initio probability of this outcome would be 2.9\% if 
all seven represent the same population;  so, in this sense,  
the difference between the two classes is significant with 
a 97\% confidence.  Two aspects of this statement are worth noting.  
(1) Given only seven objects, 97\% is the highest confidence level 
that can be attained without additional information.\footnote{   
 The Kolmogorov-Smirnov test is inappropriate here for two 
 reasons:  The $D1$ values of sets LBV1 and LBV2 do not overlap, 
 and the sample size is too small.  The $t$ test for two finite 
 samples with unequal variance formally gives a confidence 
 level of about 99\% for the distinction between LBV1 and LBV2, 
 but this entails extra assumptions that are too complicated 
to review here.  A  detail in our reasoning is that  
 $D1$ is expected to be smaller in set LBV1, as observed, and 
 not larger.   The main point is that elaborate 
 statistical analyses are not worthwhile with such a small sample 
 size.  For comments about statistical testing in general, see \\https://asaip.psu.edu/Articles/beware-the-kolmogorov-smirnov-test/. } 
2) In terms of standards for judging scientific evidence,  the above 
 result is useful {\it because it was not known at the time when the 
 two classes of LBVs were defined.\/}

\begin{figure}[!h] 
\figurenum{5}
\epsscale{0.6}
\plotone{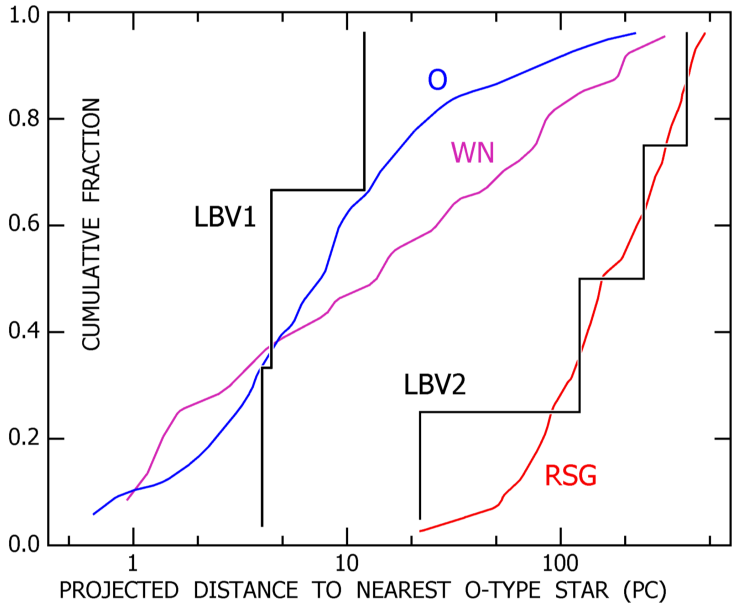}
\caption{The cumulative fraction  versus the projected distance to the nearest O
-type star for the {\it classical} LBVs (LBV1) and {\it less luminous} LBVs 
(LBV2) in the Magellanic Clouds.}
\end{figure}

Just as important, their velocities do not suggest that any
of the LBVs are runaways or have exceptionally high velocities.  
Their heliocentric velocities are included in Table 3 with the references.   
The emission line  velocities are given for the RAVE measurements \citep{Munari} except for R143\footnote{The absorption line velocities are also given in \citet{Munari}}. The published average velocities for 
 the Magellanic Clouds range   from  262 - 278 km s$^{-1}$ for the LMC 
and from 146 - 160 km s$^{-1}$ for the SMC  \citep{Richter87,McC12}.    
The LBV velocities are all consistent with their membership, 
and for the LMC stars, with their position based on the H I 
rotation curve available at NED (from  \citet{Kim}). The only exception is the candidate LBV, S119
with a possible systemic velocity based on the center of its expanding nebula 
\citep{Weis2003a} that is much lower than expected at its position in the LMC.  
Thus S119 is a candidate for a runaway star. 

In addition, a runaway star with a high velocity would be expected to alter the 
morphology of its cicumstellar ejecta, the LBV nebula. Runaway stars are
commonly known to have an arc-like  bow shock nebula due to interaction with the 
interstellar medium \citep{Bomans}. The stars in M31 and M33 are too
distant to search for bow shocks, but several confirmed LBVs and candidates in the 
Milky Way and LMC have associated circumstellar nebulae that are either spherical
or bipolar in shape \citep{Weis2011}. They are not bow shocks. The only nebula
that has a signature, an embedded arc in the surrounding nebulosity, that could be attributed to a bow shock
is S119 which may be a runaway.  

The distribution of the four LBV candidates in the LMC (Figure 6) interestingly 
follows that of the less luminous LBVs. Three of the four are considered 
candidates because of their circumstellar nebulae. The origin of their very
extended ejecta may be a prior giant eruption or numerous earlier S Dor-like 
events, but their low expansion velocities (14--27 km s$^{-1}$) are a puzzle
compared with the outflow velocities of LBV winds and giant eruptions\footnote{The nebulae associated with Milky Way LBVs have expansion velocities of 50 - 75 km s$^{-1}$.}. They may be 
due to deceleration with the ISM or the product of mass loss in a previous state 
with lower ejection velocities.  Their published luminosities are very
similar and except for Sk -69\arcdeg 279, they lie in or near the LBV 
instability strip.
Based on their position on the HR Diagram, they 
may be post-RSGs or given their rather high luminosities, former warm 
hypergiants like IRC~+10420 or Var A in M33. 

\begin{figure}[!h] 
\figurenum{6}
\epsscale{0.6}
\plotone{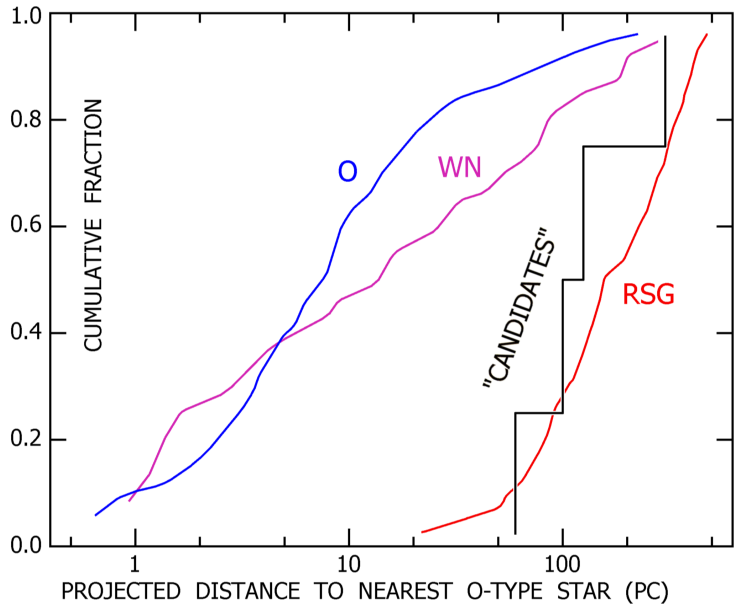}
\caption{The cumulative fraction  versus the projected distance to the nearest O
-type star for the four candidate LBVs in the LMC.}
\end{figure}

\section{Concluding Remarks}

In M31 and M33 we conclude that except possibly for 
AE And which may be a post-RSG, the LBVs and candidates are associated with luminous young stars and
supergiants appropriate to their luminosities and positions on the HR diagram.  
Their measured velocities are also consistent with their positions in their 
respective galaxies. There no evidence that they are runaway stars that have moved away from  their place of origin.   

In the Magellanic Clouds, separating the LBVs by luminosity removes the apparent 
isolation of the LBVs claimed by Smith and Tombleson. The more luminous and more massive, classical LBVs
have a  distribution similar to the late-type O stars and the 
WN stars, while the less luminous ones have a distribution like the RSGs, an evolved  lower mass population of supergiants.
In both the Magellanic Clouds and in M31 and M33, none of the sample of
16 confirmed LBVs have high velocities or are candidate runaway stars.  

One positive result is especially notable:  since the LBV distribution  
in Figure 5 is indistinguishable from the red supergiants, evidently 
the less luminous LBVs are not young.   This fact strongly supports 
the interpretation outlined in Section 1, i.e., those objects 
have evolved back to the blue in the HR Diagram. 

It is not necessary to invoke an exotic explanation such as ``kicked mass
gainers'' from  binary systems to explain the Luminous Blue Variables. 
Some LBVs and candidates are indeed binaries \citep{Rivinius,Lobel,Martayan}, but this 
binarity contradicts  
the Smith and Tombleson scenario of the LBV, the primary in 
these
systems, as the mass gainer. Thus the results of our discussion and
analysis supports  the accepted description of  LBVs of all luminosities as  evolved massive stars that have shed a lot of mass, whether as hot stars or 
 as red supergiants, and are now close to their Eddington limit.

\acknowledgements
Research by R. Humphreys, K. Davidson, and M. Gordon on massive stars is supported by  
the National Science Foundation AST-1109394.  

\appendix
\section{The LBV Environments in M31 and M33}

 The blue and H$\alpha$ images of the LBVs and candidate LBVs shown here are  from the Local Group  Galaxy Survey \citep{Massey06}. Their survey used the 
standard UBVRI filters plus narrow band filters with the Mosaic CCD Camera on
the KPNO 4-meter Mayall telescope. The reduced images have a scale of 0.27\arcsec per pixel. Details of the reductions can be found in the paper and at the LGGS web site. The authors quote 1\% to 2\% photometry for stars expected to be 
above 20M$_{\odot}$. The dashed grid on the images reproduced here is 10\arcsec. At the distances of M31 and M33, 10\arcsec is approximately 37 pc. 

\begin{figure}[!h] 
\figurenum{A1}
\plotone{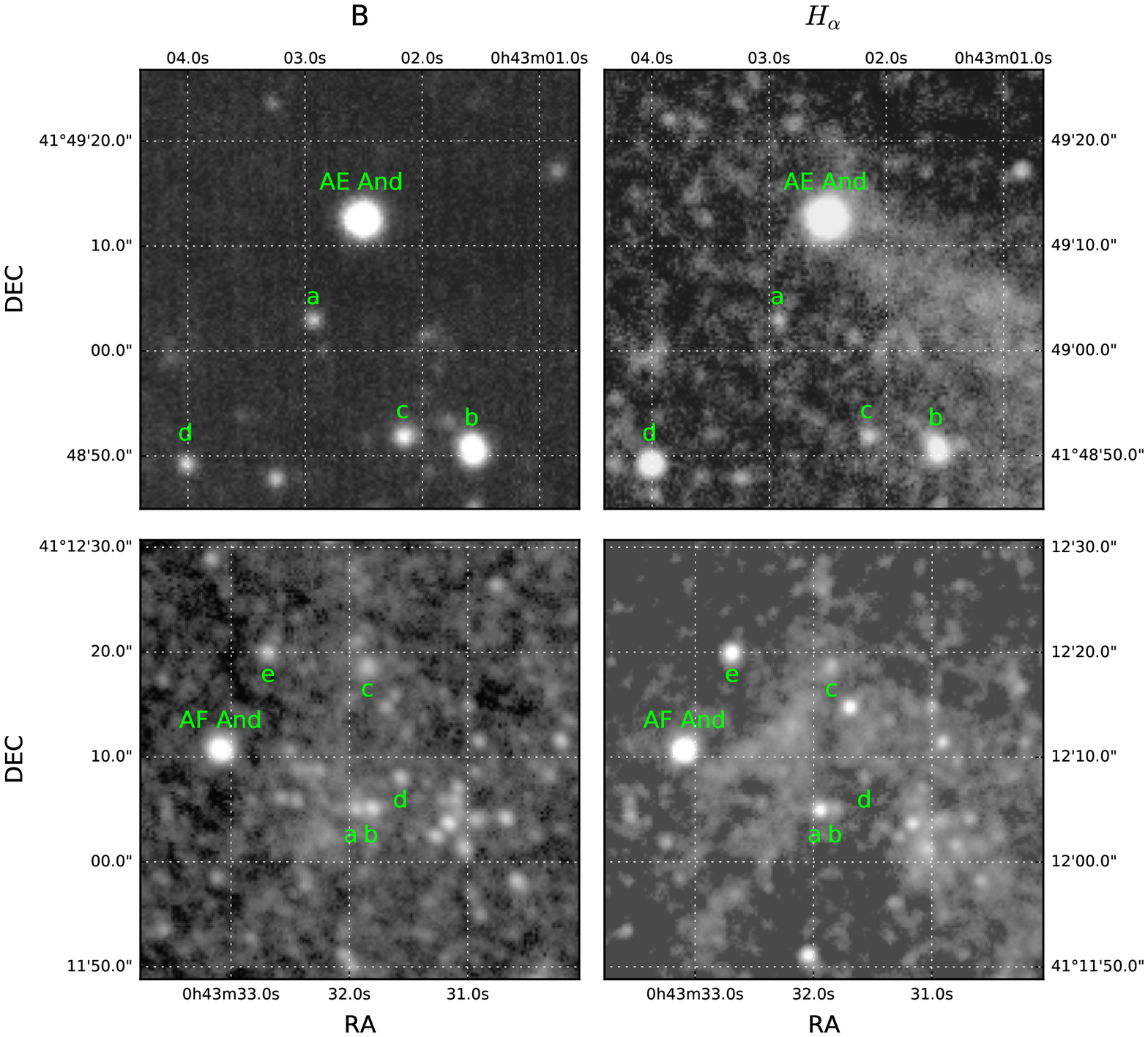}
\caption{The blue and H$\alpha$ images for AE And (top) and AF And (bottom). Note the arc of emission nebulosity near AE And.}
\end{figure}

\begin{figure}[!h]  
\figurenum{A2}
\plotone{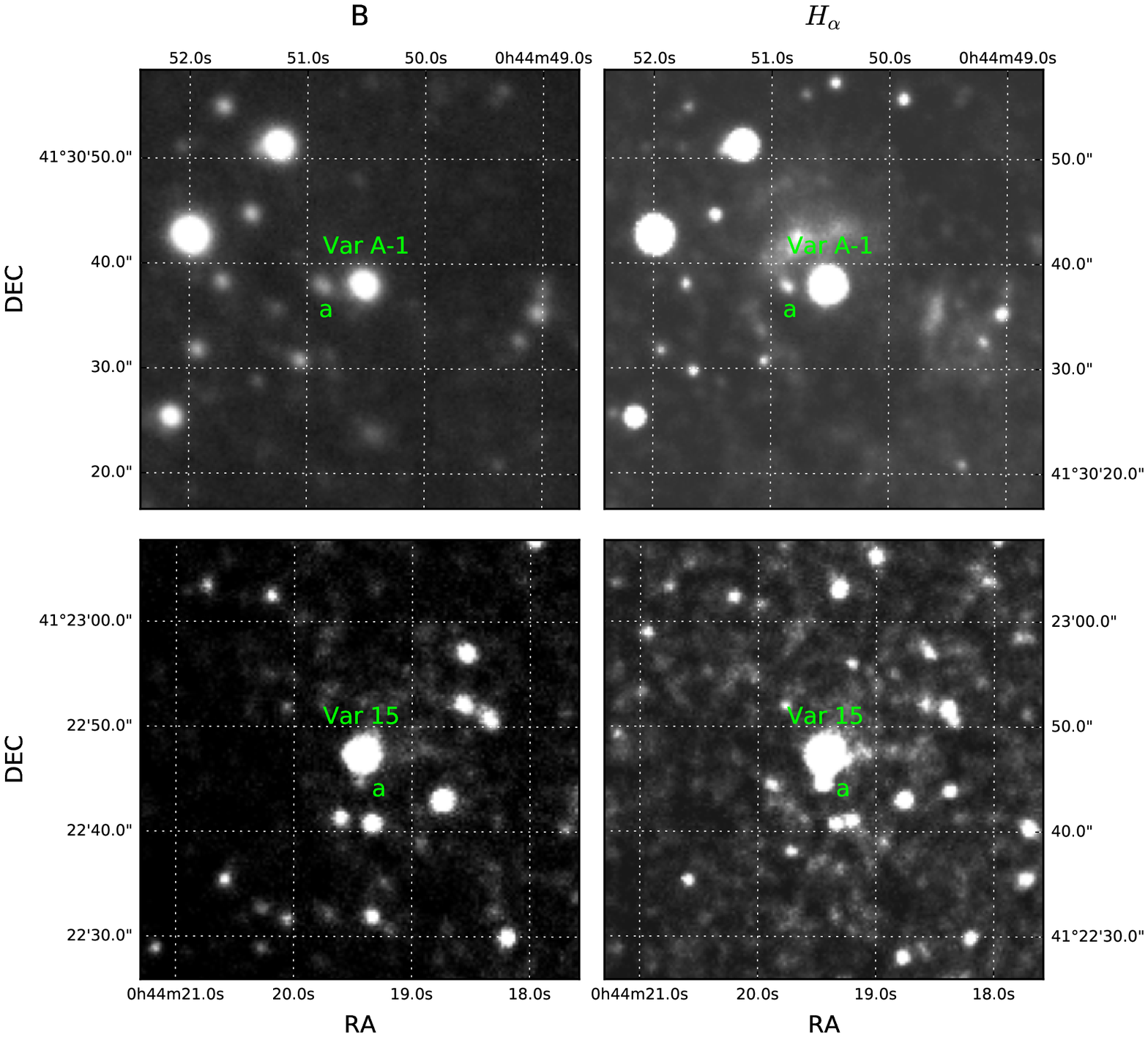}
\caption{The blue and H$\alpha$ images for Var A-1 and Var 15.}
\end{figure}

\begin{figure}[!h]  
\figurenum{A3}
\plotone{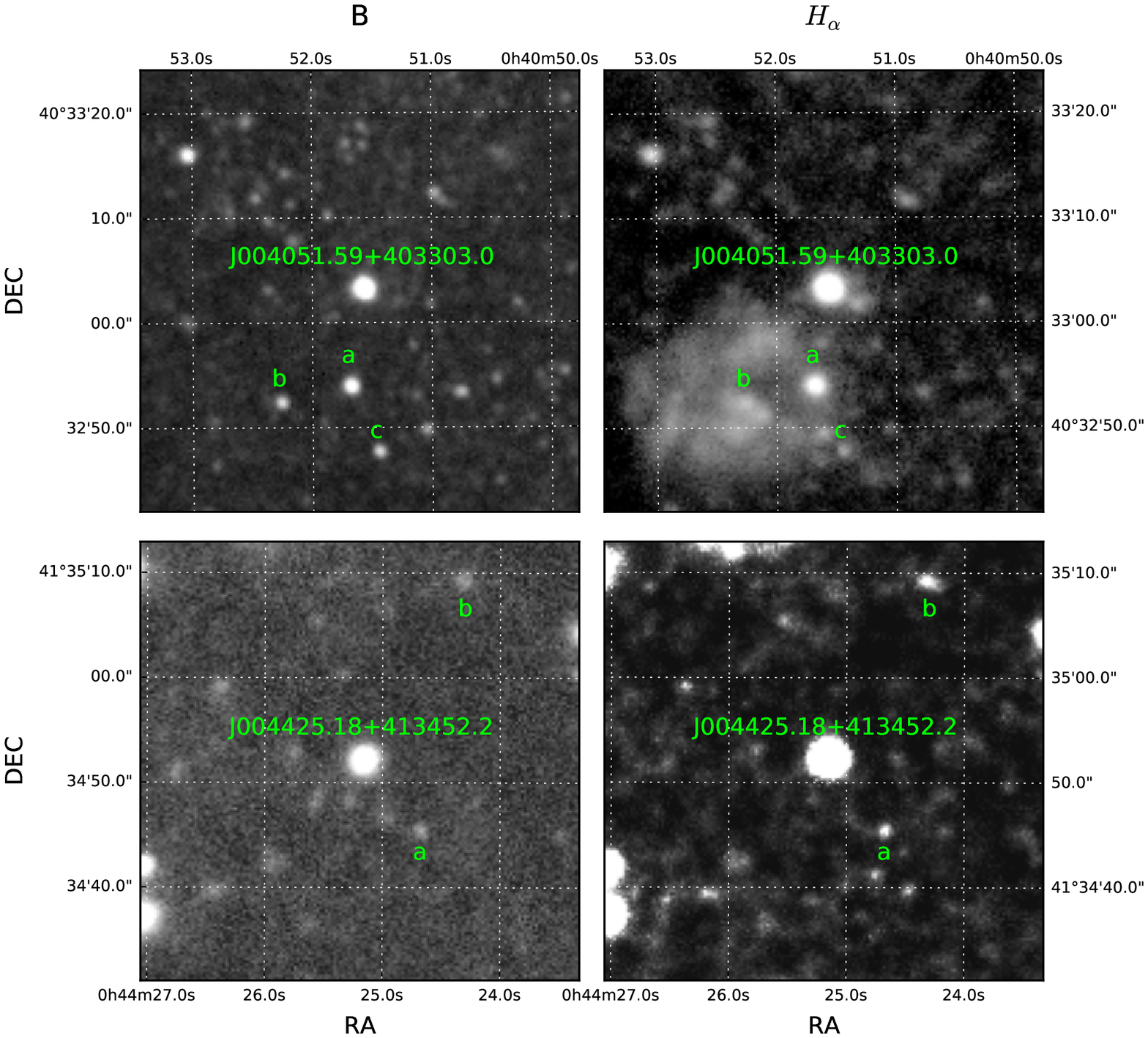}
\caption{The blue and H$\alpha$ images for J004051.59+403303.0 and J004425.18+413452.2}
\end{figure}

\begin{figure}[!h]   
\figurenum{A4}
\plotone{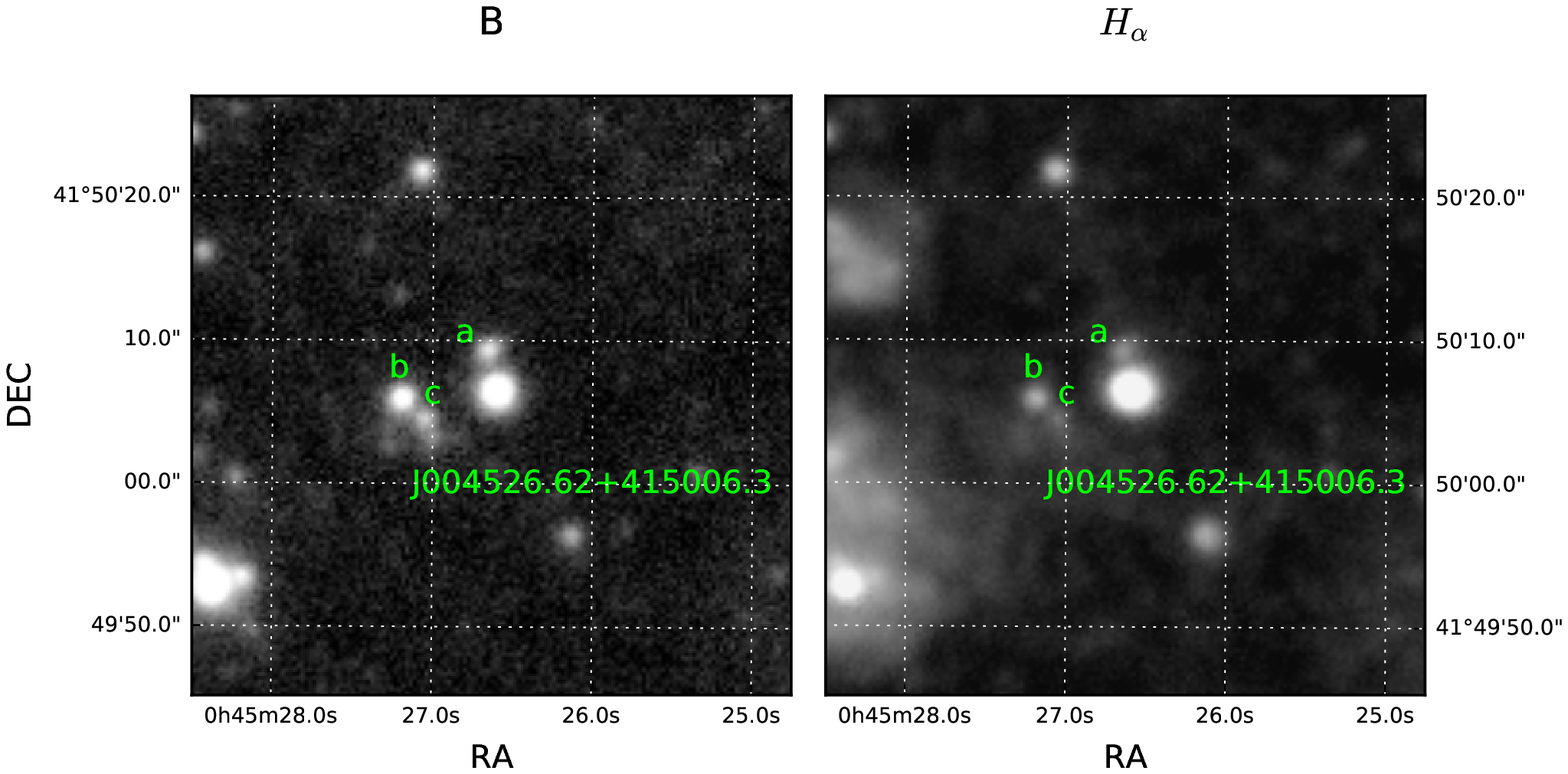}
\caption{The blue and H$\alpha$ images for J004526.62+415006.3.}
\end{figure}

\begin{figure}[!h]   
\figurenum{A5}
\plotone{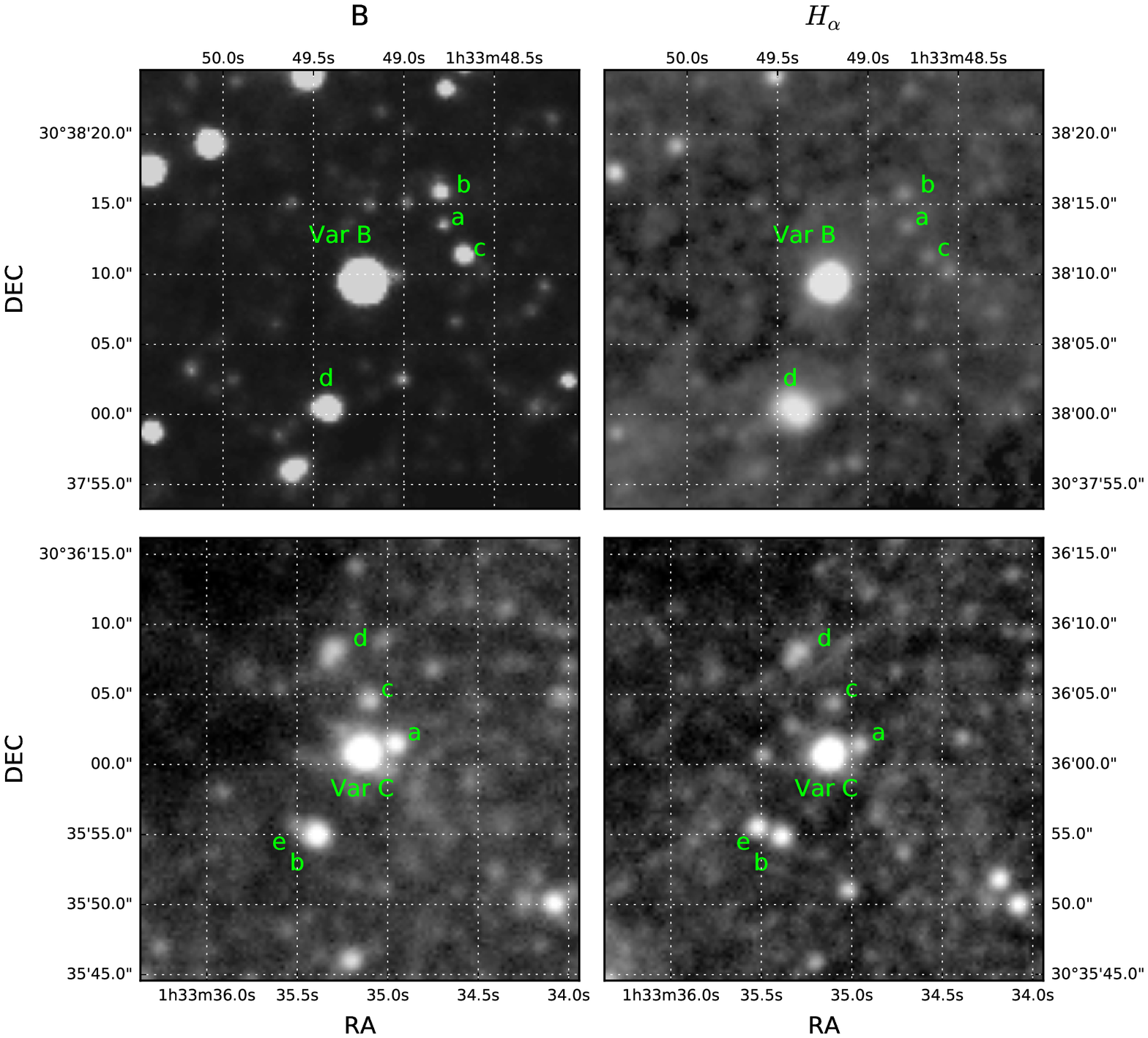}
\caption{The blue and H$\alpha$ images for Var B and Var C.}
\end{figure}

\begin{figure}[!h]   
\figurenum{A6}
\plotone{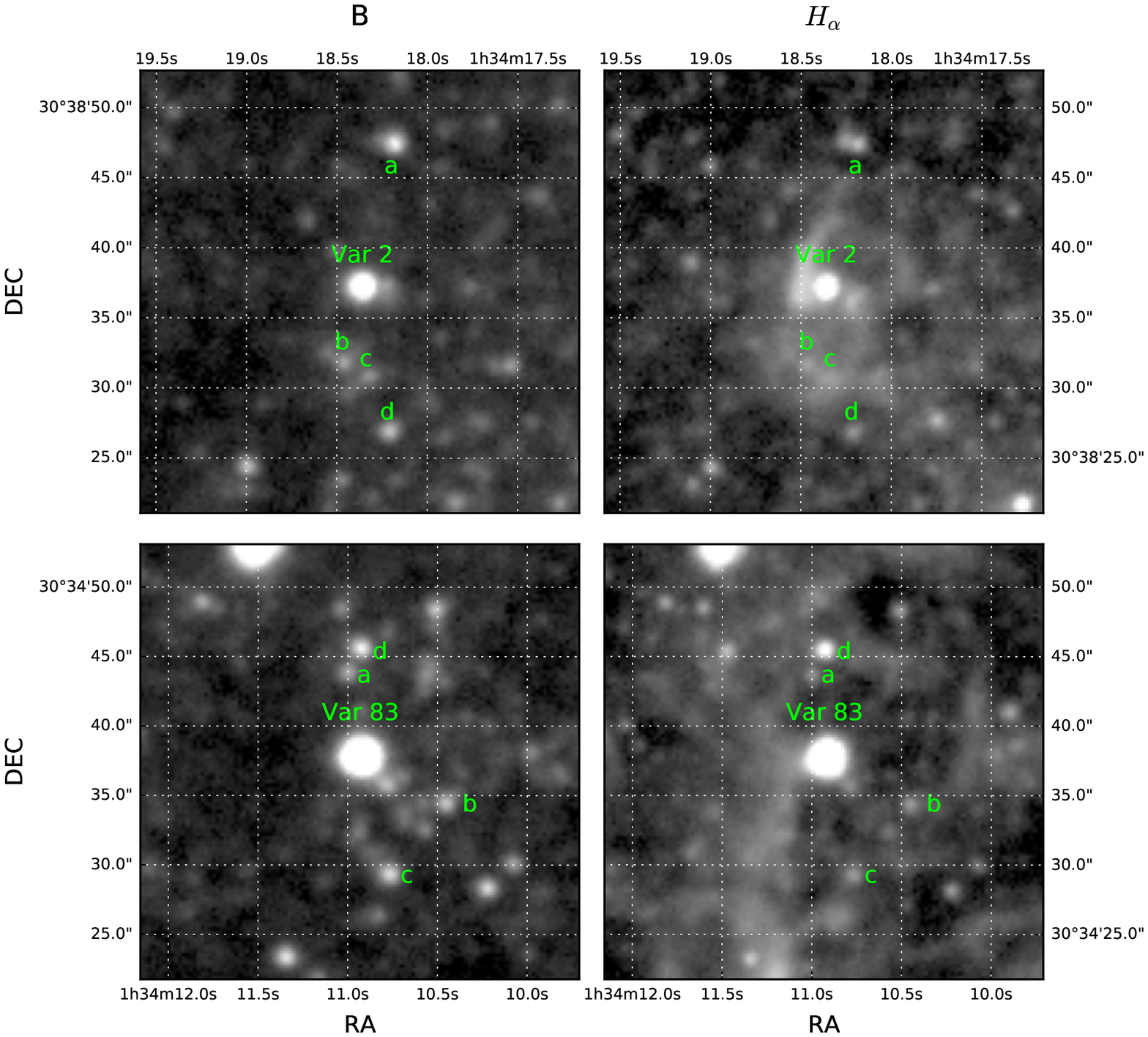}
\caption{The blue and H$\alpha$ images for Var 2 and Var 83.}
\end{figure}

\begin{figure}[!h]   
\figurenum{A7}
\plotone{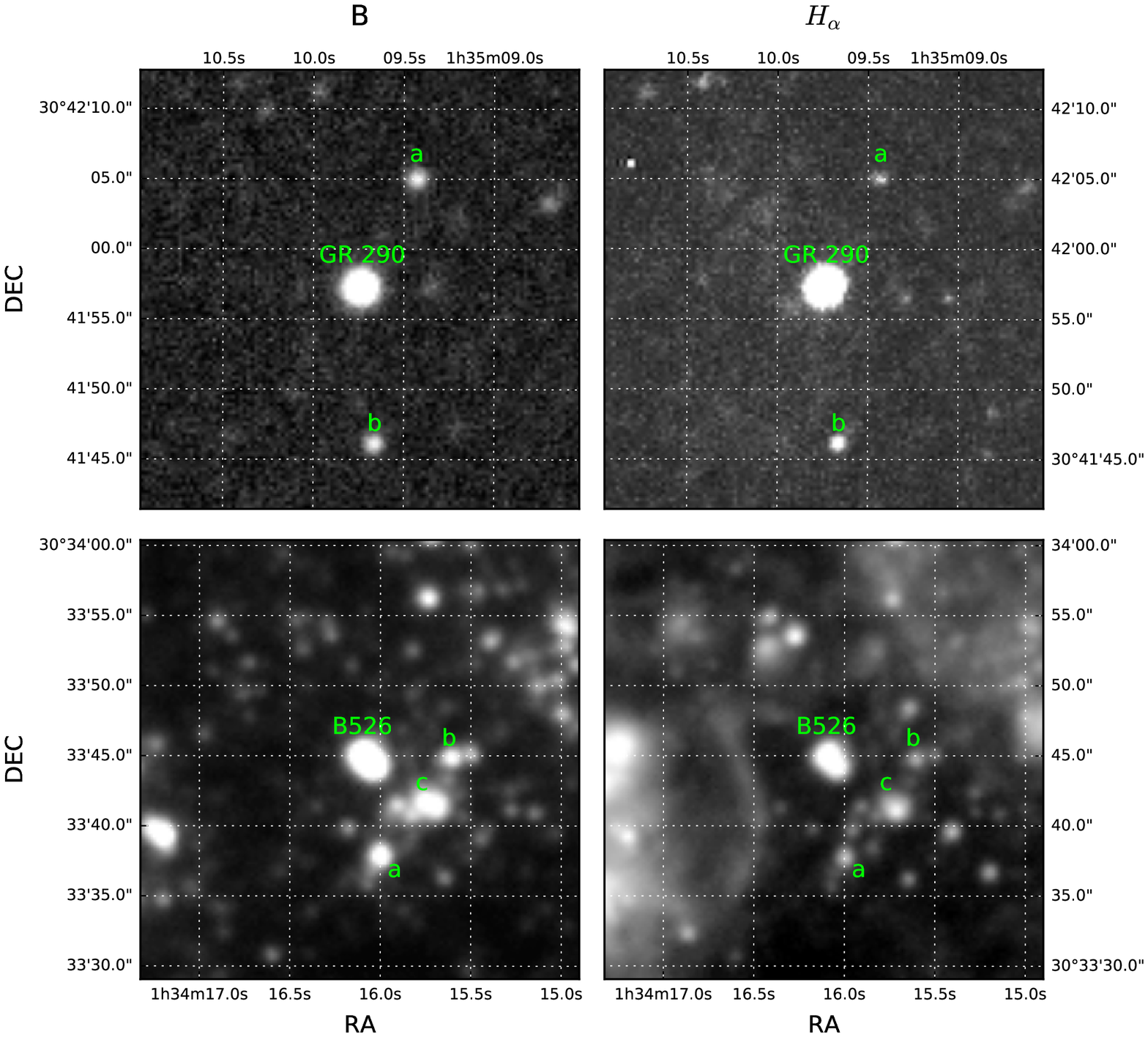}
\caption{The blue and H$\alpha$ images for GR 290 = M33-V532 and B526.}
\end{figure}

\end{document}

%% file: Table1.tex
\begin{deluxetable}{llccccl}
\tablewidth{0 pt}
\rotate
\tabletypesize{\footnotesize}
\tablenum{1} 
\tablecaption{LBVs and Candidate LBVs in M31 and M33}
\tablehead{
\colhead{Star Name}  &  \colhead{Sp. Type} &  \colhead{Quiescent Temp.} &  \colhead{M$_{Bol}$} &  \colhead{Refs\tablenotemark{a}}   & \colhead{Velocity (km s$^{-1}$)}  &   \colhead{Comment}  
}
\startdata 
   &    &   M31 &  &    &    \\
AE And  & B2-B3 &  18,000 &  -9.3 to -9.8 & 1,2 &  -250(-213)  &   A170, H II \\
AF And  & Of/WN  &  28,000 &   -10.7       & 1,2 & -274(-280)   &   \nodata \\
Var A-1 & \nodata  &   19,000 &   -10.3:      &  2  & -207(-170)   &   A42, H II \\
Var 15  &  Of/WN  &  16,000: &   -8.5:       &  2  & -243(-212)  &  A38    \\
J004526.62 +415006.3 & \nodata &  \nodata  &  -9.65 &  2,3 &  -88(-62)\tablenotemark{b}& A45, H II \\
J004051.59 +403303.0 &  \nodata  & \nodata &  \nodata  & 4  &  \nodata\tablenotemark{c}  & A82, candidate \\
J004425.18 +413452.2 &  B2-B3 &  18,000  &  -8.0     &  2  & -89(-84)\tablenotemark{d}  & A4, A9, candidate \\
       &    &   M33 &  &  &    \\
Var B   &  B0-B2  & 22,000  &  -10.4  &  1,2  & -147(-110)\tablenotemark{e}   &   A142 \\
Var C   &  B1-B2  & 20,000  &  -9.8   &  2,5    & -121(-112)  &    H II \\
Var 83  &  \nodata  & 32,000:  &  -11.1  &  1,2    & -165(-161)  &     A101, A103, H II\\ 
Var 2   &  Ofpe/WN9  &  29,500  &  -10.1  &  2,6      & -208(-193)\tablenotemark{f}   &   A100 \\
V532  &  WN8  & 42,000: &  -10.4 to -10.7 & 7  & -214(-208)\tablenotemark{g}   & A89 candidate \\
UIT 008 &  Of/WN  &  33,000  &  -10.7   &   2   & \nodata  &    A27, NGC 288, candidate \\
B526NE  &  \nodata &  \nodata &   \nodata &  2   &  -184(-163)\tablenotemark{h}   &  A101, candidate  \\
\enddata
\tablenotetext{a}{(1)\citet{Szeif},(2)\citet{RMH14},(3)\citet{RMH15},(4)\citet{Shol},(5)\citet{RMHVarC},(6)\citet{Neugent11},(7)\citet{Shol2011}}  
\tablenotetext{b}{The star was in eruption. Only two [Fe II] lines.}
\tablenotetext{c}{No [Fe II] lines.}
\tablenotetext{d}{Absorption lines.}
\tablenotetext{e}{No [Fe II] lines in the 2010 spectrum. The velocity is from the [N II] a
nd the [Fe III] lines. Two [Fe II]lines in the 2013 spectrum give -142 km s$^{-1}$. }
\tablenotetext{f}{No [Fe II] lines. The velocity is from [N II].}  
\tablenotetext{g}{No [Fe II] lines. The velocity is from the [N II] and the [Fe III] lines
.}
\tablenotetext{h}{Its close neighbor to the SW has a velocity of -179.} 
\end{deluxetable}

%% file: Table2.tex
\begin{deluxetable}{lcccccl}
\tablewidth{0 pt}
\tabletypesize{\scriptsize}
\tablenum{2}
\tablecaption{M31 and M33 LBV Neighbors}
\tablehead{
\colhead{Star Name} &   \colhead{V(mag)}  &  \colhead{B$-$V (mag)}    &
 \colhead{U$-$B (mag)}  & \colhead{(B$-$V)$_{0}$ (mag)}  & \colhead{M$_{V}$ (mag)} &  \colhead{Comment}   
 }
 \startdata
       &  M31      &        &     &     &      &      \\
AE And  &  (J004302.52+414912.4)  &      &        &     &      &      \\
a J004302.93+414902.8  &  21.04 & -0.05  & -1.07 &  -0.34 & -4.3     &      \\
b J004301.58+414850.1  &  19.59 & -0.01  &  -1.03 & -0.34   & -5.9     &      \\
c J004302.17+414851.8  &  20.53 & -0.21  &  -1.08 & -0.31  & 4.2     &      \\
d J004304.02+414849.1  &  19.49 & 1.57  & \nodata  & \nodata  &      &  RSG?  \\ 
AF And  &  (J004333.09+411210.4)     &        &        &     &      &      \\
J004331.95+411204.8 &  19.89  &  1.00 & -0.55 &  \nodata  & \nodata &  YSG? \\
J004331.82+411204.9 &  20.27 & -0.04 & -0.86  &  -0.27 & -4.9     &       \\
J004331.85+411218.4 &  20.28 &  0.62 &  0.34  &  \nodata & \nodata & YSG?      \\
J004331.57+411207.8 &  20.85 &  0.09 &  -0.79 & -0.28  &  -4.7 &       \\
J004332.69+411219.7 &  19.39 &  1.51 &  1.18  &  \nodata  & \nodata & RSG?  \\  
Var A-1 & J004450.54+413037.7)   &        &        &     &      &      \\
a J004450.89+413037.6 & 19.99 &   0.15 &  -0.81 &  -0.30 &  -5.9  &      \\ 
Var 15  & (J004419.43+412247.0)  &        &        &     &      &      \\
a J004419.45+412244.3 &  20.86 & 1.48  &  \nodata  & \nodata & \nodata & RSG? \\ 
J004526.62+415006.3 &      &        &        &     &      &      \\
a J004526.68+415009.1 & 19.49 & 0.18  & -0.80  &  -0.31 &  -6.5   &      \\
b J004527.21+415005.7 &  18.78 & -0.03 & -0.99  &  -0.32 & -6.6  &      \\
c J004527.08+415004.3 &  20.08 & -0.01 & -0.93  &  -0.31 & -5.3  &      \\
J004051.59+403303.0 &      &        &        &     &      &   candidate \\
a J004051.70+403253.8 & 18.80 & 0.58 &  -0.57  & -0.33   &  \nodata &  YSG?  \\ 
b J004052.28+403252.2 & 20.52 &  0.11 &  -1.00  & -0.35  & -5.4  &      \\
c J004051.47+403247.6 & 20.59 & 0.11  & -0.41 &  -0.16   & -4.7   &        \\ 
J004425.18+413452.2 &      &        &        &     &      &   candidate \\
a J004424.70+413445.3 & 20.69 & 0.64  &  0.33  & \nodata & \nodata & YSG  \\ 
b J004424.35+413509.1 & 20.80 & 0.39  &  0.14  & \nodata & \nodata & YSG \\ 
      &  M33      &        &     &     &      &      \\
Var B  & (J013349.23+303809.1)  &   &        &     &      &      \\
a J013348.82+303815.5 &  19.78 &  0.35 & -0.49 &  \nodata  & \nodata &      \\
b J013348.81+303816.7 &  21.70 & -0.21 & -0.63 &  \nodata & \nodata &      \\
c J013348.80+303813.2 &  20.12 &  0.68 &  0.35 &  \nodata & \nodata & YSG     \\
d J013349.43+303800.1 &  19.14 & -0.17 & 0.28  &  \nodata & \nodata & compact HII? \\
Var C   &  (J013335.14+303600.4)  &     &        &     &      &      \\
a J013334.97+303601.1 &  19.31 & -0.20 &  -1.16 & -0.33 & -5.6  &       \\
b J013335.40+303554.7 &  18.48 &  0.10 &  0.05  &  \nodata & \nodata & YSG? \\
c J013335.11+303604.2 &  20.09 &  -0.12 &  -1.11 & -0.34 & -5.1 &       \\
d J013335.30+303607.9 &  19.86 &   0.45 & -0.38  & \nodata & \nodata & YSG? \\
d J013335.53+303555.3 &  19.71 &  1.79  & 1.34   & \nodata & \nodata & RSG?  \\
Var 2 & (J013418.36+303836.9)  &        &        &     &      &      \\
a J013418.20+303847.1 &  19.72 & 0.00  &  -0.07 & -0.02  & -4.8   &       \\
b J013418.47+303831.5 &  20.69  &  0.00 & -0.93   &  -0.31 & -4.8 &       \\
c J013418.34+303830.5 &  21.03 & -0.10 &  -1.09   &  -0.34 & -4.2  &      \\
d J013418.22+303826.6 &  20.43 & -0.11 & -1.04    &  -0.32 & -4.7  &       \\
Var 83  & (J013410.93+303437.6)  &    &        &     &      &      \\
a J013411.02+303443.6 &   20.63 & -0.08 & -0.78  &  -0.24  & -4.4    &       \\
b J013410.47+303434.3 &  20.23  &  0.01 &  -1.03 &  -0.34  & -5.4    &       \\
c J013410.79+303429.2 &  20.03 & -0.12 &  -0.85  &  -0.25  & -4.9    &       \\
d J013410.95+303445.3 &  19.08  &  0.77 & 0.45   & \nodata  & \nodata& YSG?  \\
M33-V532 = GR 290 & (J013509.73+304157.3)  &     &      &     &      &  candidate     \\  
a J013509.43+304204.8 &  21.25 & -0.25  &  -1.12  &  -0.31 & -3.5 &       \\
b J013509.66+304146.2 &  20.82 & 0.25 &   0.15    &  \nodata  & \nodata & YSG?\\
B526NE (M33C-7292)&  (J013416.10+303344.9)  &    &   &    &  &  candidate \\
a J013416.01+303337.6 & 18.67 & -0.20 & -0.87 & -0.24 & -6.0  &     \\
b J013415.62+303344.7 &  18.98 & -0.09 & -0.99 & -0.31 & -6.2  &    \\
c J013415.77+303341.7 &  18.89 & 0.10 &  -0.77 & -0.28 & -6.8  &     \\ 
d J013415.71+303341.0 &  18.74 &   0.51 &  -0.82 & \nodata  &  \nodata & YSG? \\
\enddata
\end{deluxetable}

%% file: Table3.tex
\begin{deluxetable}{lccccccl}
\tabletypesize{\footnotesize}
\rotate 
\tablenum{3} 
\tablecaption{LBVs and Candidate LBVs in the Magellanic Clouds}
\tablehead{
\colhead{Star Name}  & \colhead{Sp. type}  &    \colhead{Quiescent Temp.} &  \colhead{M$_{
Bol}$} &  \colhead {Refs\tablenotemark{a}} & \colhead{Dist(pc)} &  \colhead{Velocity(km s$
^{-1}$)}   &  \colhead{Comment}  
}
\startdata
    &  LMC   &      &           &        &   &  \\
S Dor  &  B I      &  20,000 -- 25,000  &  -9.8 &  1,2  & 12   &  296(13)   & N119  \\
R127 (HDE 269858f)   &  Ofpe/WN9  &  $\sim$ 30,000      & -10.5 & 1,3  &  4   & 276(13)   & N135    \\
R143   & O9.5/B1:      & 25,000:    &  -10.0  & 4  &   4.5  &249--286(13)  &  N157 \\
R71\tablenotemark{b} (HDE 269006)    &  B5:I      & 13,600  &  -9.2 & 5,6     &  387  & 196(13)   &   \\
R110 (HDE 269662)   & B6 I      & 10,250    &  -8.9  &  7  &  244  &  248(14)    & N135   \\
R85 (HDE 269321)  &  B5Iae    &  13,000:   &  -8.5  &  8   &  22  &  292(14)  &   N119  \\
S61 (Sk 67\arcdeg 266)   & WN11h  & 27,600    & -9.7   &  9 &  125  & 294(15)    &   candidate   \\
S119 (HDE 269687)  &  WN11h  &  27,000   &  -9.8 & 9  & 302 &   156(16) &  candidate  \\
Sk -69\arcdeg 142a (HDE 269582) &  WN10h    & 22,000     &  -9.7 & 9   &     78   &\nodata  &  candidate  \\
Sk -69\arcdeg 279 &  O9f  &  30,300  &  -9.7 & 10     &   60  & \nodata  &  candidate  \\
     &  SMC   &      &           &        &  \\
HD 5980  &  WN4 + ?    & 45,000--60,000:  & -11.5: &  11  &   21   & \nodata & giant eruption \\  
R40 (HD 6884)  &  B8 Ia    & 12,000  &   -9.4  & 12  &  122  & 181(14) &   \\   
\enddata
\tablenotetext{a}{(1)\citet{Stahl86},(2)\citet{Leit},(3)\citet{Stahl83},(4)\cite{Parker},(5)\citet{Wolf81},(6)\citet{Mehner},(7)\citet{Stahl90},(8)\citet{Massey2000},(9)\citet{Crowther97},(10)\citet{Conti86},(11)\citet{Koenig14},(12)\citet{Szeifert93},(13)\citet{Munari},(14)\citet{FTW},(15)\citet{Ard}, (16)\citet{Weis2003a}}
\tablenotetext{b}{The parameters derived for R71 depend on the adopted 
foreground extinction. The analysis of its previous S Dor maximum light by 
\citet{Wolf81} used the minimum  foreground value of 0.15 mag for the LMC. 
Here we use adopt A$_{V}$ of 0.37 mag, see \citet{Mehner}, which yields an
M$_{Bol}$ of -9.2 mag for its 1970-77 eruption. During its current large
eruption, R71 has actually increased in luminosity.}
\end{deluxetable}